\documentclass{article}
\usepackage{epsfig}
\usepackage{amsmath}
\usepackage{amssymb}
\newcommand{\hep}[1]{{\tt hep-ph/#1}}
\newcommand{\jhep}[3]{{\it JHEP }{\bf #1} #3 (#2)}
\newcommand{\jetp}[3]{{\it Sov.~Phys.~JETP }{\bf #1} #3 (#2)}

\newcommand{\npb}[3]{{\it Nucl.~Phys.~}{\bf B #1} #3 (#2)}

\newcommand{\plb}[3]{{\it Phys.~Lett.~}{\bf B #1} #3 (#2)}
\newcommand{\prd}[3]{{\it Phys.~Rev.~}{\bf D #1} #3 (#2)}

\newcommand{\sjnp}[3]{{\it Sov.~J.~Nucl.~Phys.~}{\bf #1} #3 (#2)}
\newcommand{\zpc}[3]{{\it Z.~Phys.~}{\bf C #1} #3 (#2)}

\begin{document}
\titlepage      
\begin{flushright}     
February 2007
\end{flushright}      
      
\vspace*{1in}      
\begin{center}      
{\Large \bf The Pomeron in the weak and strong coupling  limits \\
and  kinematical constraints }\\      
\vspace*{0.4in}      
Anna \ M. \ Sta\'sto\\      
\vspace*{0.5cm}       
{\it Physics Department, Penn State University,\\
104 Davey Laboratory, University Park PA 16802-6300, USA}\\
{and}   \\
{\it Institute of Nuclear Physics PAN, Radzikowskiego 152,      
 Krak\'ow, Poland}
 \vskip 2mm      
\end{center}      
\vspace*{1cm}      
\centerline{(\today)}      
      
\vskip1cm      
\begin{abstract}    
We show that a resummation model for the evolution kernel at small $x$ creates a bridge  
between the weak and strong couplings. The resummation model embodies DGLAP and BFKL anomalous dimensions
at leading logarithmic orders, as well as a kinematical constraint on the real emission part of the kernel. 
In the case of pure gluodynamics the strong
coupling limit of the Pomeron intercept is consistent with the exchange  of the spin-two, colorless particle. 
\end{abstract} 

\newpage
\section{Introduction}
~~The high energy limit of QCD is one of the most important aspects  of strong interactions. The abundance of  precision data from  high - energy colliders, like HERA and Tevatron, enables one to confront  and check the theoretical  predictions in QCD. 
The standard framework used to make the predictions in QCD is that of the
collinear factorization. In this approach   the hard matrix element is computed within the perturbative QCD. The collinear singularities are factorized into the parton densities which are then evolved   
using the renormalization group  equations.  These equations provide a tool for
computing  the change of the densities of quarks and gluons at the variable  scale. Although the parton distribution evolves, the total energy-momentum is  conserved. This is  reflected  by the fact that the anomalous dimensions of the DGLAP equations
vanish at $j=\omega+1=2$
\begin{equation}
2 N_f\gamma_{qg}(\omega)|_{\omega=1} + \gamma_{gg}(\omega)|_{\omega=1}  =  0 \nonumber
\end{equation}
\noindent and
\begin{equation}
 \gamma_{gq}(\omega)|_{\omega=1} + \gamma_{qq}(\omega)|_{\omega=1}  =  0 \; , \nonumber
\end{equation}
order by order in the perturbation theory. Therefore  parton distributions $F_i(x,\mu^2)$  obey the momentum sum rule
$$
\sum_i \int_0^1 \, dx \, x F_i(x,\mu^2) \; = \; 1\; ,
$$
where $x$ is the fraction of the longitudinal momentum of the hadron  carried by the parton, and $\mu$ is the   scale of the DGLAP equation. In the distribution $F_i$ 
partons are {\it on-shell} and have zero transverse momentum. This by itself 
is already an approximation, since one does not observe free, on-shell quarks and in any hadronic process the  target  partons are always off-shell.

~~An alternative approach  is  BFKL resummation \cite{BFKL}, where 
gluon emission diagrams are summed in the Regge limit $s \gg |t|$.
The amplitude computed perturbatively in this
regime has the form
$$
A(s,t) \; \sim \;  s^{\alpha_P(t)}\; ,
$$
where $\alpha_P$ is the Pomeron trajectory   given
by  the integro-differential equation
\begin{equation}
\frac{df(x,k_T)}{d\ln 1/x} \; = \; \frac{\alpha_s N_c}{\pi}\int d^2 k_T' \, K(k_T,k_T')  \,  f(x,k_T') \;.
\label{eq:bfkl0}
\end{equation}
~~In (\ref{eq:bfkl0}) $K(k_T,k_T')$ is  the BFKL kernel which depends on the transverse momenta $k_T$. 
 In the leading logarithmic approximation the Pomeron trajectory intercept equals
$$
\alpha_P(0) \,= \,1 + 4 \ln2\, \frac{N_c \alpha_s}{\pi}\; .
$$
~~The  solution to Eq.~(\ref{eq:bfkl0}) is the unintegrated gluon distribution function $f(x,k_T)$, which still contains  the information about the transverse momentum of the gluon. The
distribution $f$ is for the {\it off-shell } gluons. 

~~The next-to-leading order corrections  \cite{NLLFL,NLLCC} dominate the leading order. Consequently, several methods on the resummation
of the perturbative series were developed,  to name \cite{Altarelli:1999vw}-\cite{Altarelli:2005ni},  \cite{Ciafaloni:1998iv}-\cite{Ciafaloni:2003kd}, \cite{THORNE} and  \cite{Kwiecinski:1997ee}.
There is, as yet unfortunately,  no unique resummation procedure at small $x$.  However, the major building 
blocks are  common to all approaches, like the DGLAP anomalous dimension, kinematical constraints, running  coupling and the momentum sum rule. 

~~The unique sum of the perturbation theory, if it exists, should contain the complete
information about the anomalous dimensions. It is then legitimate to ask what is the strong coupling limit of such   a sum. While one does not yet have a complete solution, one can still  build models and explore them in  the strong coupling limit. In  \cite{Kotikov:2003fb,Kotikov:2004er} a resummation procedure  was proposed and it revealed that the intercept
 in the limit of the infinite coupling equals one. In this paper  we investigate the  details of the behavior of the BFKL eigenvalue  restricted by  the  kinematical constraint and the momentum sum rule.
We  show that  in the limit $\alpha_s \rightarrow \infty$ the intercept of the resummed Pomeron becomes exactly one. This is consistent  with  the exchange of the (possibly massless)  particle of spin two.  The model proposed here explains the strong coupling limit but also  allows to interpolate between small and large values of $\alpha_s$. The importance of the  exact kinematics for the parton distributions  in Monte Carlo generators has been already emphasized  \cite{Collins:2003fm,Collins:2004vq}. The systematic program of the  definition of the collinear factorization with exact kinematics is under way \cite{CRS}.
This paper solely focuses   on the  BFKL Pomeron. 

~~ Our results may bear some relevance  in the context of   the string/gauge duality conjecture proposed by Maldacena  \cite{Maldacena:1997re}. Recent string theory results \cite{Brower:2006ea} (see also \cite{Janik:2000pp,Janik:1999zk})  showed that  the scattering amplitudes in the  Regge regime at large N and for the vanishing beta function  exhibit a  diffusion  in the fifth dimension of  the curved ${\rm AdS}_5$ space. The kernel has  the form
\begin{equation}
K(r,r',s) \; = \; \frac{s^{j_0}}{\sqrt{4\pi D \ln s}} e^{-(\ln r-\ln r')^2/(4 D \ln s)} \; , \nonumber
\end{equation}
with 
\begin{eqnarray}
j_0 & = & 2-\frac{2}{\sqrt{\lambda}}+{\cal O}(1/\lambda)  \; , \nonumber \\
D & = & \frac{1}{2\sqrt{\lambda}}+{\cal O}(1/\lambda) \; , \nonumber
\end{eqnarray}  
where $\lambda=R^4/\alpha^{\prime 2}$,  $R$ is the radius of the ${\rm AdS}_5$ space and the  string tension equals $1/(2\pi \alpha')$. In the case of the supersymmetric Yang-Mills theory
$\lambda=g^2 N=4\pi\alpha_s N$.
Therefore the diffusion properties  of the BFKL Pomeron are universal   at all values of the strong coupling. We  show that the  resummed eigenvalue  is  consistent with the results of \cite{Brower:2006ea}.

\section{BFKL evolution and higher order corrections}

~~In the leading logarithmic 
approximation 
the BFKL evolution equation \cite{BFKL} in the 
momentum representation reads
 \begin{multline}
f(x,k_T) = f^{(0)}(x,k_T) \; + \\
\; + \, \frac{\alpha_s N_c}{\pi} \, \int_x^1\frac{dz}{z}\,\int \frac{k_T^2 dk_T^{\prime 2}}{k_T^{\prime 2}}\, \bigg\{ \frac{f(\frac{x}{z},k_T^{\prime })-f(\frac{x}{z},k_T)}{|k_T^{\prime 2}-k_T^2|}
\,+\,\frac{f(\frac{x}{z},k_T)}{[4k_T'^4+k_T^4]^{1/2}}\bigg\} \; ,
\label{eq:bfkl}
\end{multline}
where $x$ is the Bjorken variable, and $k_T,k_T'$ are  transverse momenta of the exchanged gluons.
  For the unintegrated gluon distribution function $f(x,k_T)$  the solution to the  Eq.~(\ref{eq:bfkl})  behaves as
\begin{equation}
f \sim x^{-\omega_P} \; ,
\end{equation}
where
\begin{equation}
\omega_P \, \equiv \, \alpha_P(0) -1 \, = \, 4 \ln 2 \, \frac{\alpha_s N_c}{\pi} \, = \, \chi_0(\gamma=1/2) \; ,
\end{equation}
and the kernel eigenvalue $\chi_0$  has the form
\begin{equation}
\chi_0(\gamma) \; = \; -2\gamma_E-\psi(\gamma)-\psi(1-\gamma) \; ,
\label{eq:chi0}
\end{equation}
where $\psi$ is the digamma function and $\gamma$ is the Mellin  conjugate to $\ln k_T^2$.
The next-to-leading corrections to the BFKL equation are known to be very large \cite{NLLFL,NLLCC}. The resummation procedure  yields a  stable result for the intercept. The following  
features  are common to all methods of resummation 
\begin{itemize}
\item Full DGLAP anomalous dimension (at least in the  leading order of perturbation theory).
\item Kinematical constraint imposed onto the gluon emission term (explained below).
\item Running  coupling constant.
\item Momentum sum rule.
\end{itemize}

It has been shown, \cite{Ciafaloni:1999yw} that the  Pomeron intercept computed from the renormalization group-improved BFKL equation with  modifications mentioned above,  is much smaller than  the leading logarithmic value.  The resummed value increases with the coupling constant.
The 
$\alpha_s^2$ correction which comes from the  expansion of the resummed result coincides with the
NLLx \cite{NLLFL,NLLCC} calculation down to a couple of percent.

The kinematical constraint \cite{AGS,KMS,Ciaf88} (for  Monte Carlo results within dipole picture  see \cite{Avsar:2006jy,Avsar:2005iz}) is imposed onto the BFKL equation as a requirement that the virtualities of the   exchanged gluons  are dominated  by the transverse
 components of the momenta.
The form of the  kinematical constraint depends on the scales in the scattering process \cite{Salam1998}. For
DIS reaction which is characterized by  a large difference of scales of the scattered particles (where virtuality of the photon  $Q^2$ is much larger than the scale on the proton side $\sim \Lambda_{QCD}^2$) the kinematical constraint
imposed onto the real emission term in (\ref{eq:bfkl}) approximately equals
\begin{equation}
k_T^{\prime 2} \, \le \, \frac{k_T^2}{z}  \; .
\label{eq:kca}
\end{equation}

In the case of the scattering of objects with similar scales (as for example in $\gamma^*\gamma^*$ scattering with photons having both large and comparable virtualities), the kinematical constraint has a symmetrical form
\begin{equation}
\Theta[k_T/z-k_T'] \, \Theta[k_T'/z-k_T] \; .
\label{eq:kc}
\end{equation}

 The exact  form
of the constraint, as it appeared in \cite{AGS,KMS},  actually is
\begin{equation}
q_T^2 \, \le \, k_T^2 \,\frac{1-z}{z}\; , \hspace*{1.0cm} q_T \, \equiv\, k_T-k'_T \; .
\label{eq:kca_exact}
\end{equation}
~~In the limit of small $z$  we   neglect it in the numerator of (\ref{eq:kca_exact}) and set $k_T'^2 \simeq q_T^2$.
The  comparison   between the two alternative  versions of the constraint (\ref{eq:kca}) and (\ref{eq:kca_exact})  is relegated to  the final section of this paper.

\begin{figure}[htb]
\centerline{\epsfig{file=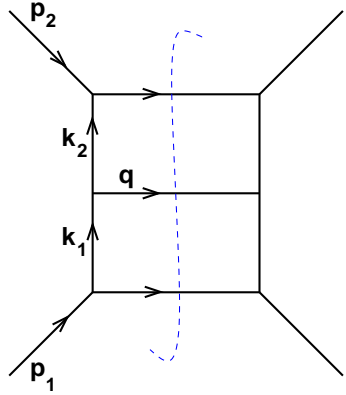,width=0.3\textwidth}}
\caption{Elementary diagram for the one particle emission in the high energy scattering.}
\label{fig:0}
\end{figure}

 More exact analysis of the   kinematics and the constraint  can be best illustrated by the following simple 
example. In Fig.~\ref{fig:0} we show a  $2\rightarrow 3$ process with one particle emission. The details  of the matrix element are irrelevant for the purpose of  this discussion. The incoming momenta of the particles satisfy the condition   $p_1^2\simeq p_2^2 \simeq m^2 \ll s$.  The general expression for the phase space of  such a diagram is    
\begin{equation}
\int \, dPS_3 \equiv \int \frac{d^4 k_1}{(2\pi)^4} \frac{d^4 k_2}{(2\pi)^4}  \, \delta((k_1-k_2)^2) \, \delta((p_1-k_1)^2) \,\delta((p_2+k_2)^2)\, .
\end{equation}
~~We decompose  the exchanged particle momenta in terms of Sudakov variables  \\
$k_i=(\alpha_i p_1^+,\beta_i p_2^-,k_{iT})$. The phase space can be rewritten as
\begin{multline}
\int \, dPS_3 \equiv   \frac{s^2}{4(2\pi)^8} \int d\alpha_1 d\beta_1 d^2 k_{1T}\,d\alpha_2 d\beta_2 d^2 k_{2T} 
\cdot \delta\left(s(\alpha_1-\alpha_2)(\beta_1-\beta_2)-(k_{1T}-k_{2T})^2\right) \\
\cdot \,
\delta(-s(1-\alpha_1)\beta_1-k_{1T}^2) \,  \delta(s(1+\beta_2)\alpha_2-k_{2T}^2) \; .
\end{multline}
~~In the limit of the multi-Regge kinematics 
$$
0 \ll \alpha_2 \ll \alpha_1 \ll 1\; , \hspace*{0.5cm} 0 \ll |\beta_1| \ll |\beta_2| \ll 1 \; ,
$$
which leads to the following approximated form for the phase space
\begin{multline}
\int \, dPS_3 \equiv   \frac{s^2}{4(2\pi)^8} \int d\alpha_1 d\alpha_2 d\beta_1  d\beta_2 \, d^2 k_{1T} \,  d^2 k_{2T} \,\\
\cdot\,  \delta\left(s \, \alpha_1 \, \beta_2 +(k_{1T}-k_{2T})^2\right)\, \delta(-s \beta_1-k_{1T}^2) \, \delta(s \alpha_2-k_{2T}^2)
  \, \; .
\end{multline}
~~Furhermore, in the high energy limit  the transverse momenta are approximated as
$k_{1T}^2 \simeq k_{2T}^2  \simeq m^2 $ which leads to 
\begin{equation}
\label{eq:psfact}
\int \, dPS_3 \equiv   \frac{1}{4(2\pi)^8} \int_{\alpha_2}^1 \frac{d\alpha_1}{\alpha_1} \, d\alpha_2 \, \delta(s \alpha_2-k_{2T}^2)  \, d^2 k_{1T} \,  d^2 k_{2T}  =  \frac{1}{4(2\pi)^8s} \ln \frac{s}{m^2} \, \int \, d^2 k_{1T} \,  d^2 k_{2T}  \; .
\end{equation}
~~Phase space becomes  completely factorized into transverse and longitudinal components. This is  also transparent  in Eq. (\ref{eq:bfkl}), where we identify $x\equiv m^2/s,\alpha_1 \equiv x/z $.  From the point of view
of the  leading logarthmic accuracy. i.e. $(\alpha_s \ln s/m^2)^n$, the choice of $m^2$ is irrelevant since it has to be much smaller than the very  large energy $s$. This, however, leads to 
a mismatch in the kinematics for the outgoing particle. The approximated  phase space 
is now completely factorized, as in Eq.~(\ref{eq:psfact}), and the $k_T$ integration is unrestricted.
This can give an unphysical result for  the emitted gluon since
it can be    off-shell 
\begin{eqnarray}
z\rightarrow 1 & \longrightarrow & q^+ \rightarrow 0 \; ,\nonumber \\
q^2 =2 q^+ q^- -q_T^2 & \simeq & - q_T^2 <0 \;  {\rm if } \;  k_T \neq k_T'\; .
\end{eqnarray}
The kinematical constraint  (\ref{eq:kca_exact}) guarantees that in the limit $z\rightarrow 1$ the emitted gluons are soft and on-shell ( $q^2 \simeq 2 q^+ q^- \simeq q_T^2 \simeq 0$). Note that  the condition (\ref{eq:kca_exact}) constrains only the phase space.  The matrix element
is kept in the same, high-energy approximation. 
\section{BFKL with kinematical constraint}

~~Let us now explore  the  details of  the effects of the symmetrical kinematical constraint (\ref{eq:kc}) imposed on the BFKL kernel.
Throughout this analysis we will keep the strong coupling fixed.
The modified equation reads\footnote{We write here $f(x,k_T^2)$ instead of $f(x,k_T)$ to emphasize that the angular dependence has been integrated out.}
\begin{multline}
f(x,k_T^2) = f^{(0)}(x,k_T^2) \; + \\
\; + \, \bar{\alpha}_s \, \int_x^1\frac{dz}{z}\,\int \frac{k_T^2 dk_T^{\prime 2}}{k_T^{\prime 2}}\bigg\{ \frac{f(\frac{x}{z},k_T^{\prime 2})\Theta(k_T'-k_Tz)\Theta(k_T/z-k_T')-f(\frac{x}{z},k_T^2)}{|k_T^{\prime 2}-k_T^2|}+\\
+\;\frac{f(\frac{x}{z},k_T^2)}{[4k_T'^4+k_T^4]^{1/2}}\bigg\} \; ,
\label{eq:bfklkc}
\end{multline}
where we   introduced  standard notation $\bar{\alpha}_s\equiv\alpha_s N_c/\pi$.
We perform the following change of the variables
\begin{eqnarray}
\eta & = & \bar{\alpha}_s \ln \frac{1}{z} \; , \nonumber \\
\xi & = & \bar{\alpha}_s  \ln \frac{1}{x} \; ,
\label{eq:change variables}
\end{eqnarray}
after which  the equation (\ref{eq:bfklkc}) can be rewritten as
\begin{multline}
f(\xi,k_T^2) = f^{(0)}(\xi,k_T^2) \; + \\
\; + \, \int_0^{\xi} d\eta\, \int \frac{k_T^2 dk_T^{\prime 2}}{k_T^{\prime 2}}\bigg\{ \frac{f(\eta-\xi,k_T^{\prime 2})\,\Theta[k'-k\exp(-\eta/\bar{\alpha}_s)]\,\Theta[k_T\exp(\eta/\bar{\alpha}_s)-k_T']-f(\eta-\xi,k_T^2)}{|\,k_T^{\prime 2}-k_T^2\,|}+\\
+\;\frac{f(\eta-\xi,k_T^2)}{[\,4k_T'^4+k_T^4\,]^{1/2}}\bigg\}\; .
\label{eq:bfklkc1}
\end{multline}
The LO BFKL is recovered when the theta functions are set to one and  its solution can be written
in a form of series in $\xi$
$$
f^{LO}(\xi,k_T^2) \;  =  \; \sum_i \xi^i \, c_i(k_T^2) \; ,
$$ 
which  is the leading logarithmic series in $\alpha_s \ln 1/x$. 
Clearly, the non-perturbative nature  of the kinematical constraint is  evident because of the appearance of the  exponential
factors $\exp(\pm\eta/\bar{\alpha}_s)$. It is non-perturbative in a sense that the leading logarithmic expansion in $\eta$ is accompanied by the arbitrary powers in $\bar{\alpha}_s$.
The resummed 
Eq.~\ref{eq:bfklkc1} contains all powers of $\bar{\alpha}_s$ as well as $\xi$ and the solution
is a  function of three variables $f(\xi,\bar{\alpha}_s,k_T^2)$, or $f(x,\bar{\alpha}_s,k_T^2)$.

 For the  fixed values of $\eta$ variable the $\Theta$ functions have the following behavior
\begin{eqnarray}
\Theta[k_T'-k_T\exp(-\frac{\eta}{\bar{\alpha}_s})]\,\Theta[k_T\exp(\frac{\eta}{\bar{\alpha}_s})-k_T'] & \stackrel{\alpha_s \rightarrow 0}{\longrightarrow}  & 1 \; , \\
\Theta[k_T'-k_T\exp(-\frac{\eta}{\bar{\alpha}_s})]\,\Theta[k\exp(\frac{\eta}{\bar{\alpha}_s})-k_T'] & \stackrel{\alpha_s \rightarrow \infty}{\longrightarrow}  & \frac{2\eta}{\bar{\alpha}_s}\delta(k_T-k_T')\; .
\label{eq:kclargealphas}
\end{eqnarray}
~~The constraint  breaks down the complete factorization between the longitudinal and transverse components of the momenta in the evolution. The evolution equation cannot be written any longer in a simple  form of the differential equation
with the factorized $\ln 1/x$ dependence as in  Eq.~\ref{eq:bfkl0}.

The Mellin transform of the kernel in
 Eq.~(\ref{eq:bfklkc}) reads

\begin{multline}
\chi(\gamma,\omega) \; = \;   \int_0^1\frac{dz}{z} z^{\omega}\,\int \frac{k_T^2 dk_T^{\prime 2}}{k_T^{\prime 2}}\bigg\{ \frac{\Theta(k_T'-k_Tz)\,\Theta(k_T/z-k') \left(\frac{k^{'2}}{k_T^2}\right)^{\gamma}-1}{|\,k_T^{ \prime 2}-k_T^2\,|}
+\;\frac{1}{[\,4k_T'^4+k_T^4\,]^{1/2}}\bigg\} =  \\  \\
=\,\int_0^1\frac{dz}{z}z^{\omega}\left[u^{\omega/2}\int_0^1\frac{du}{u}\frac{u^{\gamma}-1}{1-u}+u^{-\omega/2}\int_0^1\frac{du}{u}\frac{u^{\gamma}-1}{u-1}+\int_0^{\infty}\frac{du}{u}\frac{1}{\sqrt{4u+1}}\right] = \\ \\
 = \; \; \; -2\gamma_E-\psi(\gamma+\frac{\omega}{2})-\psi(1-\gamma+\frac{\omega}{2})\; .
\label{eq:kernel_mellin}
\end{multline}

\begin{figure}[htb]
\centerline{\epsfig{file=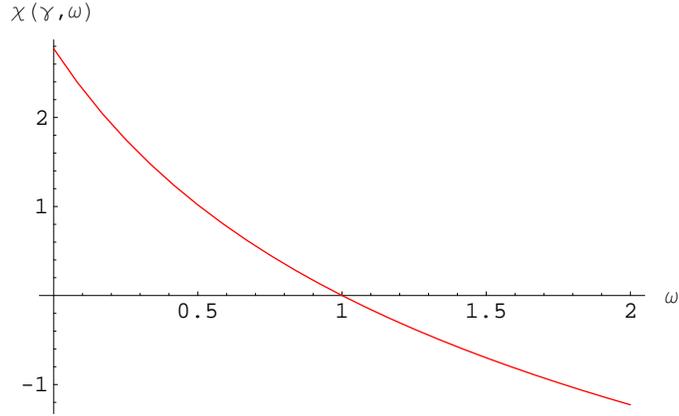,width=0.6\textwidth}}
\caption{The behaviour of the eigenvalue function $\chi(\gamma,\omega)$ , Eq.~\ref{eq:kernel_mellin}, at $\gamma=1/2$ and around $\omega=1$. }
\label{fig:1}
\end{figure}
\begin{figure}[htb]
\centerline{\epsfig{file=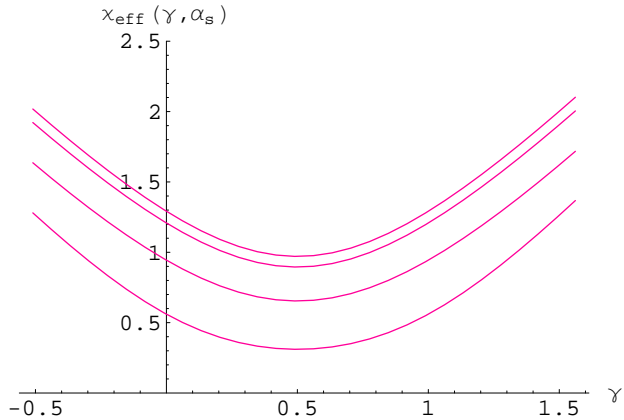,width=0.6\textwidth}}
\caption{The solution to the eigenvalue equation (\ref{eq:chieffkc}) in the case of the LO BFKL, with kinematical constraint  in  the form of the function $\chi^{\rm eff}(\gamma,\alpha_s)$, for four different values of the strong coupling: $\bar{\alpha}_s=0.2,\,1.0,\,5.0,\,20.0$, from the  bottom to the top curve. }
\label{fig:2}
\end{figure}

The effect of the kinematical constraint is to shift the arguments of the digamma  functions.
The arguments $\gamma+\omega/2$ and $1-\gamma+\omega/2$ can now be equal or  larger than $1$ in the domain $0<\gamma<1, \, \omega>0$. Therefore,  both BFKL branches, $\psi(1)-\psi(\gamma+\omega/2)$ and $\psi(1)-\psi(1-\gamma+\omega/2)$,   have zeros in this regime.
This results in a  solution $\chi\left(\gamma_0,\omega_0(\gamma_0)\right)=0$ with $0<\gamma_0<1$. In particular, the 
 eigenvalue $\chi(\gamma,\omega)$ (\ref{eq:kernel_mellin}) vanishes when $\gamma\rightarrow 1/2$ and $\omega\rightarrow 1$
\begin{equation}
 \chi(\gamma=1/2,\omega=1) \; = \; 0 \; ,
\label{eq:limit_chi}
\end{equation}
which is illustrated in Fig.~\ref{fig:1}. The  solution to the eigenvalue equation
\begin{equation}
\omega \; = \; \bar{\alpha}_s\, \chi(\gamma,\omega) \; ,
\label{eq:chiomgamma}
\end{equation}
can be  symbolically written as \cite{Ciafaloni:2003rd} 
\begin{equation}
\omega \; = \; \chi^{\rm eff}(\gamma,\bar{\alpha}_s) \; .
\label{eq:chieffkc}
\end{equation}
~~It is difficult to obtain the solution analytically, however we can   easily solve numerically for $\chi^{\rm eff}$.  In Fig.~\ref{fig:2} we illustrate
this solution  as a function of $\gamma$ for different values of $\bar{\alpha}_s$. For increasing values of $\bar{\alpha}_s$, the minimum of $\chi^{ \rm eff}$ at $\gamma=1/2$ approaches the  limit of $1$ as is expected from Eq.~\ref{eq:limit_chi}. 
The shape of the function changes very little with the increase of  $\bar{\alpha}_s$.
The derivative of $\chi(\gamma,\omega)$ at $\omega=1$ equals  
\begin{equation}
\frac{d\chi(\gamma=1/2,\omega)}{d\omega}|_{\omega=1} \; = \; -\frac{\pi^2}{6} \; .
\end{equation}
~~Therefore, we can write an expansion of the eigenvalue (\ref{eq:chiomgamma})
around $\omega=1$
\begin{equation}
\chi(\gamma=1/2,\omega) \; \simeq  \; (1-\omega) \frac{\pi^2}{6} \; ,
\label{eq:chiomexpand}
\end{equation}
however, the  shifted kernel eigenvalue is  not sufficient to reproduce the full NLLx BFKL result.
~~Motivated by the resummation procedure proposed in \cite{Ciafaloni:1999yw,Ciafaloni:2003rd} let us  consider the resummed model for the eigenvalue equation in the form
\begin{equation}
\frac{1}{\bar{\alpha}_s} \; = \; \bigg[\frac{1}{\omega}+A(\omega)\bigg] \, \chi(\gamma,\omega) \; .
\label{eq:bfklresum_lo}
\end{equation}
The expression in the square brackets is the LO DGLAP anomalous dimension
\begin{equation}
\gamma^{(0)}(\omega) \; = \; \frac{1}{\omega} + A(\omega) \; ,
\end{equation}

\begin{equation}
A(\omega)\; =\; -\frac{1}{\omega+1}+\frac{1}{\omega+2}-\frac{1}{\omega+3}-\psi(2+\omega)-\gamma_E
+\frac{11}{12}\; ,
\end{equation}
with 
\begin{equation}
A(0)\; =\; -\frac{11}{12} \; ,
\end{equation}
and 
\begin{equation}
\gamma^{(0)} \, = \, 0\; \;\; \leftrightarrow \; \; \; A(1) \, = \,  -1 \; .
\label{eq:sumrule}
\end{equation}
~~The last condition, expressed by formulae (\ref{eq:sumrule}), is the momentum sum rule in the absence of quarks $N_f=0$.
Therefore, the resummed model (\ref{eq:bfklresum_lo}) contains  DGLAP anomalous dimension at to leading order in $\alpha_s \ln Q^2$, 
BFKL eigenvalue at  the leading order in $\ln 1/x$ and the kinematical constraint. We  again stress, that the coupling constant is fixed in this analysis.
In particular, this enables us to use simple expressions for the eigenvalue conditions like Eq.~(\ref{eq:chiomgamma}).
We  expand the DGLAP anomalous dimension $\gamma^{(0)}(\omega)$ around $\omega=1$ and  get
\begin{equation}
\gamma^{(0)} \; \simeq \; (1-\omega) \bigg(\frac{\pi^2}{6}-\frac{65}{144}\bigg) \; = \; 1.1935 \,(1-\omega) \; .
\label{eq:anomexpand}
\end{equation}
~~Using (\ref{eq:chiomexpand}) and (\ref{eq:anomexpand}) we can thus write (\ref{eq:bfklresum_lo})
\begin{equation}
\frac{1}{\bar{\alpha}_s} \; = \; (1-\omega)^2 \,\frac{\pi^2}{6} \,\bigg(\frac{\pi^2}{6}-\frac{65}{144}\bigg)\;  =\; 1.9633\, (1-\omega)^2  \; .
\label{eq:largealphas}
\end{equation}
~~Solving for $\omega$ we obtain
\begin{equation}
\omega\; =\; 1-\frac{c_0}{\sqrt{\bar{\alpha}_s}}  \; ,
\label{eq:omegalargealphas}
\end{equation}
where
\begin{equation}
c_0 \; = \;  1/\sqrt{\frac{\pi^2}{6}\bigg(\frac{\pi^2}{6}-\frac{65}{144}\bigg)}  \; =\;  0.509346 \; .
\end{equation}
~~The result (\ref{eq:omegalargealphas}) should be compared with the calculation  of the first correction to the  intercept of the graviton presented  in \cite{Brower:2006ea},
$$
j=\,1+\omega\,=\,2-\frac{c_0}{\sqrt{\bar{\alpha}_s}}, \hspace*{1cm} c_0 = 1/\pi \; ,
$$
 and the  SUSY result  \cite{Kotikov:2004er}. 
The coefficient $c_0$ obtained from the model (\ref{eq:bfklresum_lo})  is different,  due to different details of the resummation model and the fact that here we are considering fixed coupling  QCD, whereas Refs.~\cite{Brower:2006ea,Kotikov:2004er} apply to 
 the SUSY YM case.  However, the overall agreement is  satisfactory.
\begin{figure}[htb]
\centerline{\epsfig{file=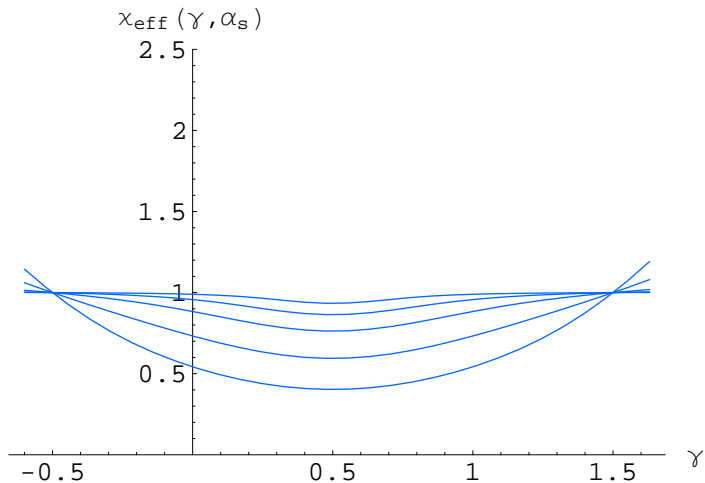,width=0.7\textwidth}}
\caption{The solution to the eigenvalue equation (\ref{eq:chieffkc}) in the case of the LO BFKL with kinematical constraint and the DGLAP anomalous dimension in  the form of the function $\tilde{\chi}^{\rm eff}(\gamma,\alpha_s)$ for the  five different values of the strong coupling $\bar{\alpha}_s=0.2,\,1.0,\,5.0,\,20.0,\,100.0$ from the  bottom to the top curve. }
\label{fig:3}
\end{figure}

The model (\ref{eq:bfklresum_lo}) provides  not only the correct limits at small and large values of the coupling but also smoothly  interpolates between these limiting regions.
From (\ref{eq:largealphas}) it is clear that the $1/\sqrt{\bar{\alpha}_s}$ behaviour is a results of  the double zero of
the model (\ref{eq:bfklresum_lo}) at $\omega=1$.  The numerical solution to the equation (\ref{eq:bfklresum_lo}) in the form of \footnote{We put an additional  tilde $\tilde{}$ to distinguish it from the result  of  Eq.~\ref{eq:chieffkc}.}
$$
\omega=\tilde{\chi}^{\rm eff}(\gamma,\alpha_s) \; ,
$$
 is plotted in Fig.\ref{fig:3} where we show  it for various values of the coupling constant.
The fixed points in the  $(\gamma,\omega)$ plane, namely $(-1/2,1)$ and $(3/2,1)$, result from the superposition of the zero $\omega=1$ and the simple poles at $\gamma+\omega/2$ and $1-\gamma+\omega/2$. The second derivative $\frac{d^2\tilde{\chi}^{\rm eff}(\gamma,\alpha_s)}{d\gamma^2}|_{\gamma=1/2}$  appreciably changes in this case, exhibiting $1/\alpha_s$ behaviour at large $\alpha_s$. This is different from  the AdS/CFT correspondence result
which as well produces $1/\sqrt{\alpha_s}$ for the second derivative.  We think that this is an artifact of the simplified model which
we are considering. The overall qualitative behavior of $\tilde{\chi}^{\rm eff}$ is nonetheless consistent with the expectation from
string theory side, in particular when we compare with the  Fig.~2 in Ref.~\cite{Brower:2006ea}. The effective eigenvalue has two fixed points
(the points which do not change when the coupling is varied) and it can be approximated by  the parabola with a minimum at $\gamma=1/2$. When the coupling is increased, the second derivative decreases. This continues until the limit of $\alpha_s = \infty$ is reached, and the effective eigenvalue tends to  a  constant equal to one.

~~When we compare the  solutions presented in Fig.~\ref{fig:2} and  Fig.~\ref{fig:3}, we see a qualitatively different behavior of the effective eigenvalue function. In Fig.~\ref{fig:2} the
function $\chi^{\rm eff}$ is 1 only at a single point in $\gamma=1/2$. In Fig.~\ref{fig:3}
the effective eigenvalue function $\tilde{\chi}^{\rm eff}$ has a constant limit when $\bar{\alpha}_s \rightarrow \infty$ for all values of $\gamma$, which is the consequence of the fact that
$1+\omega A(\omega)\sim (1-\omega)$ at $\omega=1$.

\begin{figure}[htb]
\centerline{\epsfig{file=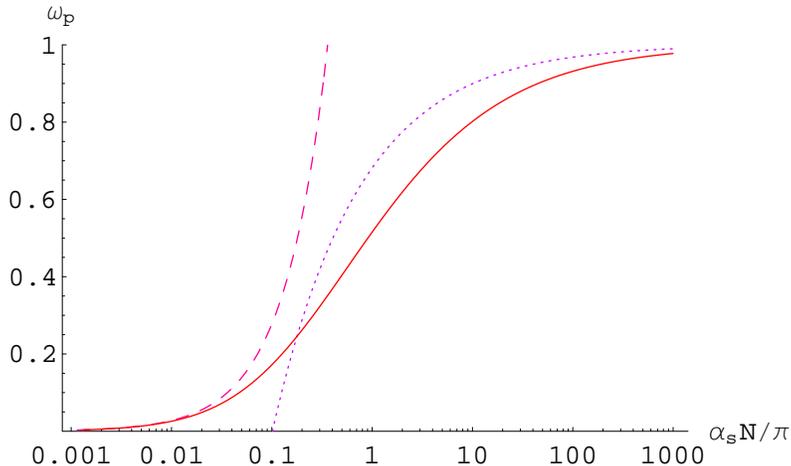,width=0.7\textwidth}}
\caption{The strong coupling dependence of the Pomeron intercept in the case of fixed coupling.
Comparison of the predictions of the resummed model with LO DGLAP (\ref{eq:bfklresum_lo}) (solid line) to the first order results in QCD \cite{BFKL} (dashed  line at small couplings ) and strings in curved background (dotted line at large coupling)  \cite{Brower:2006ea}. Note the logarithmic horizontal axis for the coupling $\bar{\alpha}_s=\frac{\alpha_s N_c}{\pi}$. }
\label{fig:4}
\end{figure}

From $\tilde{\chi}^{\rm eff}(\gamma,\bar{\alpha}_s)$ one can compute  the dependence of the Pomeron intercept 
$\omega_P=\tilde{\chi}^{\rm eff}(1/2,\bar{\alpha}_s) \; ,$
 on the coupling $\bar{\alpha}_s$  which we illustrate in Fig.~\ref{fig:4} compared with the first order expansions around $\bar{\alpha}_s\sim 0$ (leading logarithmic BFKL \cite{BFKL}) and $\bar{\alpha}_s \sim \infty$ (the string theory in the curved background \cite{Brower:2006ea}). 

It is also  interesting to examine the convergence properties of the resummed model at small and large values of $\bar{\alpha}_s$.
The coefficients of the expansion for the intercept $\omega_P=\tilde{\chi}^{\rm eff}(1/2,\bar{\alpha}_s)$ are collected in Table~1. The  left column contains the first five coefficients of the series for the Pomeron intercept 
at small values of $\bar{\alpha}_s$ $$\omega_P=\sum_{n \ge 1} \, a_n \, \bar{\alpha}_s^n \; ,$$ and the right column contains the first five coefficients of the series at large values of $\alpha_s$  $$\omega_P=1-\sum_{n \ge 1} \, b_n \, \frac{1}{{\bar{\alpha}_s}^{n/2}} \; .$$  The coefficients in the small coupling expansion grow with increasing order as $a_{n+1}/a_n \simeq -10$ signalling the possible lack of convergence of this model at $\bar{\alpha}_s \rightarrow 0$. At large values of the coupling the coefficients behave at least 
as $b_{n+1}/b_n \sim 1/n$. 
\vspace*{0.3cm}

\begin{table}[htb]
\begin{center}
\begin{tabular}{|c|c|}
\hline
$a_n$&$b_n$\\
\hline
2.77&0.713685\\
-20.72&0.300767\\
226.69&0.085747\\
-3000.73&0.015120\\
44267.30&0.000863\\
\hline
\end{tabular}
\caption{Numerical values of the expansion coefficients of the intercept at small (left column) and large (right column)
values of the strong coupling constant $\bar{\alpha}_s$.}
\end{center}
\end{table}

\section{ Improved kinematical constraint}

It is interesting to investigate the changes to the evolution kernel when imposing more stringent kinematical constraint.
In \cite{KMS} it was shown that the more accurate version of the  kinematical constraint has actually the form
\begin{equation}
q_T^2 < k_T^2 \, \frac{1-z}{z}\; .
\label{eq:kcanew}
\end{equation}
~~It reduces to Eq.~(\ref{eq:kca}) when $z \ll 1$. Let us first consider the collinear approximation to the BFKL kernel 
\begin{equation}
f(x,k_T^2) = f^{(0)}(x,k_T^2) \, + \, \int_x^1 \frac{dz}{z} \left[ \int^{k_T^2} \frac{dk_T^{\prime 2}}{k_T^{\prime 2}}+\int_{k_T^2} \frac{k_T^2 dk_T^{\prime 2}}{k_T'^4} \right] f(x/z,k_T^{\prime 2}) \; ,
\label{eq:collinear}
\end{equation}
and keep the $(1-z)$ term in the numerator of the kinematical constraint replacing
however $q_T^2 \simeq k_T'^2$.
 This constraint  also divides  the domains of the integration over the longitudinal variable $z$. There are three distinct regions. For large values of $z>1/2$, $(1-z)/z<1$ and $k_T^{\prime 2}<k_T^2$  we have
\begin{multline}
\chi^{coll,l}_{z>1/2}(\gamma,\omega) \;=\;\int_0^1 \frac{dz}{z} z^{\omega} \int^{k_T^2}_0 \frac{dk_T^{\prime 2}}{k_T^{\prime 2}} \left(\frac{k_T^{\prime 2}}{k_T^2}\right)^{\gamma} \, \Theta(k_T^2(1-z)/z-k_T^{\prime 2})\, \Theta(z-1/2)\;= \\
=\;\frac{1}{\gamma} \, \beta_{1/2}(1+\gamma,\omega-\gamma)\;=\;\frac{1}{\gamma(1+\gamma)} \; {}_2F_1(1+\gamma,1+\omega;2+\gamma;-1) \; ,
\end{multline}
and for the region where $k_T^{\prime 2}>k_T^2$ and $z<1/2,\; (1-z)/z>1$ we obtain
\begin{multline}
{\chi}^{coll,u}_{z<1/2}(\gamma,\omega) \;= \;\int_0^1 \frac{dz}{z} z^{\omega} \int^{\infty}_{k_T^2}\frac{k_T^2dk_T^{\prime 2}}{k_T'^4} \left(\frac{k_T^{\prime 2}}{k_T^2}\right)^{\gamma} \, \Theta(k_T^2(1-z)/z-k_T^{\prime 2}) \,\Theta(1/2-z) = \\
=\; \frac{1}{\omega} \, \beta_{1/2}(1-\gamma+\omega,\gamma-1)\;=\;\frac{1}{\omega(1-\gamma+\omega)} \; {}_2F_1(1-\gamma+\omega,-\omega;2-\gamma+\omega;-1) \; ,
\end{multline}
and finally $z<1/2, k_T'<k_T$ 
$$
{\chi}^{coll,l}_{z<1/2}(\gamma,\omega)=\int_0^1 \frac{dz}{z} z^{\omega} \int^{k_T^2}_0 \frac{dk_T^{\prime 2}}{k_T^{\prime 2}} \left(\frac{k_T^{\prime 2}}{k_T^2}\right)^{\gamma} \,  \Theta(1/2-z) = \frac{2^{-\omega}}{\gamma\omega}\; .
$$
~~The eigenvalue equation for (\ref{eq:collinear}) is then
$$
1=\bar{\alpha}_s \left({\chi}^{coll,l}_{z>1/2}(\gamma,\omega)+ {\chi}^{coll,u}_{z<1/2}(\gamma,\omega)+ {\chi}^{coll,l}_{z<1/2}(\gamma,\omega)\right)  \; .
$$
~~The effect of the improved kinematical constraint on the collinear model   results in the shift of the pole at $\gamma=1 \rightarrow \gamma=1+\omega$. Additionaly the regularized beta functions generate the  additional  poles at $\gamma=-1,-2,-3,\dots$ from ${\chi}_{z>1/2}(\gamma,\omega)$ and $\gamma=2+\omega,3+\omega,\dots$ from ${\chi}_{z<1/2}(\gamma,\omega)$. These singularities are located in  the same positions as those in the BFKL eigenvalue (\ref{eq:kernel_mellin}) with an asymmetric shift of the $\gamma$ poles.  This has to be compared with the following expression
\begin{equation}
\frac{1}{\omega}\, \chi^{coll}(\gamma,\omega) \; = \; \frac{1}{\omega} \left( \frac{1}{\gamma}+\frac{1}{1-\gamma+\omega}\right) \; , 
\label{eq:collinearshift}
\end{equation}
in which the 
small-z version of the constraint $k'^2<k^2/z$ imposed onto the collinear model leads to the shift of one of the existing poles in $\gamma$ space but does not introduce any new  singularities.

\begin{figure}[htb]
\centerline{\epsfig{file=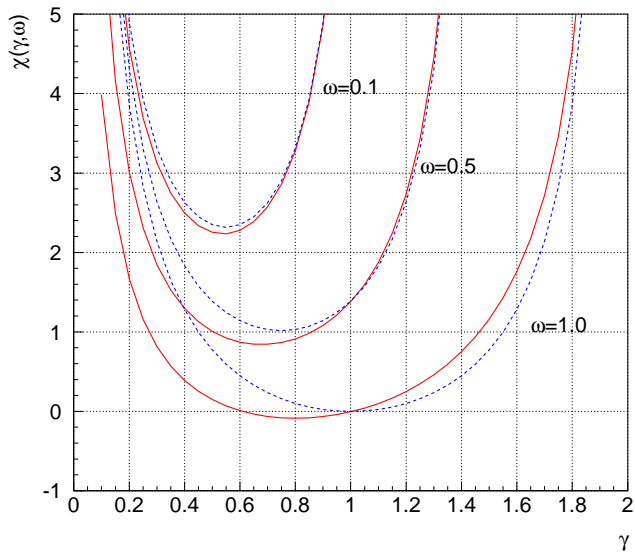,width=0.6\textwidth}}
\caption{ The eigenvalue of the BFKL kernel  with the kinematical constraint of the form
(\ref{eq:kca}) dashed lines, and (\ref{eq:kca_exact}) solid lines. Calculation performed for three different values of  the Mellin variable $\omega=0.1,0.5,1.0$. }
\label{fig:5}
\end{figure}

We  next impose the  constraint  $q_T^2 < k_T^2 (1-z)/z$ in the BFKL case
\begin{multline}
f(x,k_T) \; = \; f^{(0)}(x,k_T) \, + \, \\
+ \; \bar{\alpha}_s \, \int_x^1 \frac{dz}{z} \int \frac{d^2k_T'}{q_T^2} \left(\frac{k_T^2}{k_T^{\prime 2}}\, f(\frac{x}{z},k_T')\,\Theta\left[k_T^2 \frac{1-z}{z}-q_T^2\right]\, - \, \frac{k_T^2}{k_T^{\prime 2}+q_T^2}\, f(\frac{x}{z},k_T) \right)\; .
\end{multline}
~~In this case we perform the diagonalization of the kernel numerically. 
The eigenvalue $\chi(\gamma,\omega)$ as a function of $\gamma$ for different values of $\omega$ is shown in Fig.~\ref{fig:5}, where it is   compared with the expression (\ref{eq:kernel_mellin}) for the case of the LLx BFKL with the approximated kinematical constraint.
The  eigenvalue is further lowered  with respect to (\ref{eq:kernel_mellin}) but the overall differences are   moderate. The most affected region is the one in the vicinity of  $\gamma \rightarrow 0$. This  has to be understood as an effect of the large value of  $z$ in the cutoff $(1-z)/z <1$ for $z>1/2$, and therefore the cutoff becomes more stringent  than the previous one in the 
collinear region where $k_T'^2<k_T^2$.

%
\section{Conclusions}
~~The BFKL equation in the leading logarithmic approximation suffers from
the problem of the violation of the energy - momentum conservation.
The kinematical constraint reduces the phase space allowed for the the real emissions
and is responsible for the  resummation of the subleading corrections.
The  kernel eigenvalue in that case  is zero when $\omega=1$ and $\gamma=1/2$. This 
means that in the limit $\bar{\alpha}_s\rightarrow \infty$ the intercept becomes equal to 1, which can be interpreted as an  exchange of the color- singlet  object  with spin 2. 
When the momentum sum rule is additionally  imposed, the effective eigenvalue becomes constant and equal to  one in the limit of the  infinite strong  coupling. Such model then can provide an interpolation between the small and large values of the strong coupling constant. The first correction to the intercept of the graviton, of the form $ 1/\sqrt{\bar{\alpha}_s}$  can be also computed from this model. The more accurate form of the kinematical constraint
further reduces the eigenvalue. However, the overall qualitative behavior remains unchaged  and the model  still has to be supplemented by the requirement that at $\omega=1$ the eigenvalue vanishes for all values of the Mellin variable $\gamma$.

This result is valid in the case  of one Pomeron exchange. We stress that the result for the  intercept will be renormalized when the multi-Pomeron exchanges or the  saturation corrections  are also incorporated.  These corrections will certainly become important in the limit of the  very large coupling.  Also, the running of the coupling and the presence of  quarks in the evolution  can further change the results
in a  quantitative way.  We leave these questions for further studies.
\section*{Acknowledgments}
Discussions with Marcello Ciafaloni, Dimitri Colferai, John Collins, Leszek Motyka,  Radu Roiban,  Ted Rogers and Gavin Salam are kindly acknowledged.
This research has been supported by the Polish Committee for Scientific Research
grant No. KBN 1 P03B 028 28.

\end{document}